\documentclass[]{spie}

\usepackage{amsmath,amsfonts,amssymb}
\usepackage{graphicx}
\usepackage{booktabs}
\usepackage{tabularx}
\usepackage{array}
\usepackage[colorlinks=true,allcolors=blue]{hyperref}

\newcolumntype{Y}{>{\raggedright\arraybackslash}X}

\title{On-sky demonstration of dual-field interferometry at the CHARA Array}

\author{Narsireddy Anugu\supit{a}, 
John D. Monnier\supit{b}, 
Douglas R. Gies \supit{a},
Jeremy Jones\supit{a}, 
Robert Klement \supit{a},
Stefan Kraus\supit{c},
Rainer K\"{o}hler\supit{a}, 
Karolina Kubiak\supit{a},
Cyprien Lanthermann\supit{a},
Edgar R. Ligon\supit{a},
Denis Mourard\supit{d},   
Gail H. Schaefer\supit{a}, 
Nicholas J. Scott\supit{a}
\skiplinehalf
\supit{a}The CHARA Array of Georgia State University, Mount Wilson Observatory,
Mount Wilson, California, USA;\\
\supit{b}Department of Astronomy, University of Michigan, Ann Arbor, Michigan, USA;\\
\supit{c}School of Physics and Astronomy, University of Exeter, Exeter, United Kingdom\\
\supit{d}Universit\'e C\^ote d'Azur, Observatoire de la C\^ote d'Azur, CNRS,
Laboratoire Lagrange, Nice, France;\\
}

\authorinfo{Further author information: N.A.: E-mail: nanugu@gsu.edu}
\begin{document}
\maketitle

\begin{abstract}
Dual-field interferometry uses a bright reference star for real-time fringe tracking, allowing a second beam combiner to record long coherent integrations on a fainter off-axis science target. At the Center for High Angular Resolution Astronomy (CHARA) Array, we implement this mode using the six-telescope MIRC-X and MYSTIC beam combiners in the $H$ and $K$ bands, respectively. We first demonstrated this capability in summer 2025 on the hierarchical triple $\alpha$~Piscium. MIRC-X tracked component A in the $H$ band, while MYSTIC observed component B in the $K$ band, resolving the 7~mas Ba--Bb subsystem and measuring the relative astrometry of the 1.85~arcsec A--B pair with an uncertainty of 234~$\mu$as. Here, we describe subsequent phase-tracking testing, preliminary
sensitivity simulations, and planned instrumental upgrades aimed at
extending this mode to faint off-axis science targets.
\end{abstract}


\section{INTRODUCTION}
\label{sec:introduction}

The sensitivity of classical optical interferometry is generally limited by the short exposure time required to freeze atmospheric piston. In the near-infrared, this is typically set by the atmospheric coherence time of approximately 20--40~ms. Consequently, routine near-infrared observations with the CHARA Array \cite{TenBrummelaar2005} reach limiting magnitudes of approximately $H$ or $K\simeq8$--9 under favorable conditions \cite{Anugu2020MIRCX,Setterholm2023MYSTIC,Lanthermann2024}.

Dual-field interferometry can overcome this limitation by using a nearby bright star as a phase reference, allowing longer coherent integrations on a fainter science target. Light from two stars within the interferometric field is separated and propagated through two optical channels. A bright reference source is observed at high frame rate by a fringe tracker, whose delay corrections are applied to the common CHARA delay lines. A second beam combiner can then integrate coherently on the fainter off-axis science source. The differential optical path between the two fields contains both the projected sky separation and internal instrumental terms, allowing the same architecture to provide precise relative astrometry. The Very Large Telescope Interferometer (VLTI)/GRAVITY instrument has demonstrated the scientific potential of this mode in the $K$ band \cite{Gravity2017,Lacour2019FT}. Phase-referenced observations with GRAVITY have enabled Galactic-center astrometry and studies of faint companions, including the first direct interferometric detection and spectroscopy of the exoplanet HR~8799~e \cite{Gravity2019HR8799}. These results motivate similar faint-companion observations at CHARA, although its smaller 1~m apertures place it in a different sensitivity regime.

The initial CHARA dual-field ``first light" commissioning results on $\alpha$~Psc were presented by Anugu et al. \cite{Anugu2026AlphaPsc}. In that experiment, MIRC-X\cite{Anugu2020MIRCX} tracked component A in the $H$ band while MYSTIC\cite{Setterholm2023MYSTIC}  recorded the $K$-band science fringes from component B. The observations directly resolved the previously known spectroscopic Ba--Bb subsystem and measured the differential astrometry of the wide A--B pair. This established that dual-field interferometry is feasible at CHARA. However, the two wide components were comparably bright, and the experiment therefore did not determine the limiting magnitude of a faint science channel.

In this report, we describe the work now underway to convert this initial demonstration into a repeatable observing mode for faint off-axis science targets.

\begin{figure}[t]
\begin{center}
\includegraphics[width=0.98\textwidth]{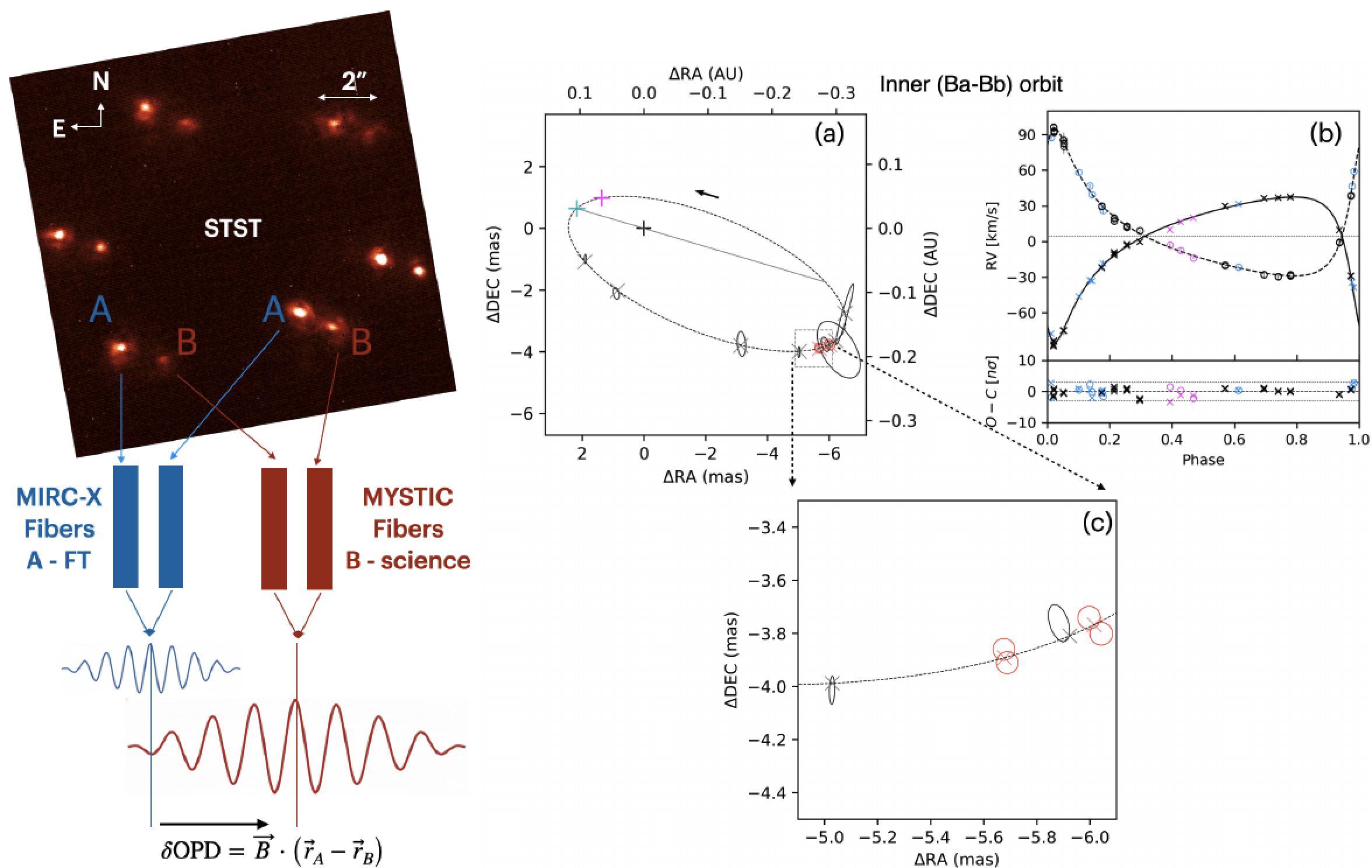}
\end{center}
\caption{First CHARA dual-field observations of $\alpha$~Psc.  Left: STST images both wide components; component A feeds the MIRC-X fringe-tracking channel and component B feeds the MYSTIC science channel.  Right: MYSTIC resolves and constrains the inner Ba--Bb orbit.  Adapted from Anugu et al.\cite{Anugu2026AlphaPsc}}
\label{fig:alpha_psc}
\end{figure}

\section{FROM THE FIRST DEMONSTRATION TO FAINT-TARGET OPERATION}
\label{sec:demonstration}

The $\alpha$~Psc observations provided an end-to-end test of the dual-star observing sequence. The Six Telescope Star Tracker (STST)\cite{Setterholm2023MYSTIC} simultaneously imaged components A and B, separated by approximately 1.85~arcsec, and provided the field information required to place each component on the appropriate set of single-mode fibers. MIRC-X received component A and supplied the real-time fringe corrections. MYSTIC received component B through the second fiber set and differential delay lines. 

Figure~\ref{fig:alpha_psc} summarizes the observing geometry and the resulting inner orbit. MYSTIC resolved the Ba--Bb subsystem at a projected separation of approximately 7~mas. Combining the interferometric measurements with radial velocities produced a 25-day orbit and dynamical masses for the two nearly identical F-type stars \cite{Anugu2026AlphaPsc}. The dual-field delay measurements also provided A--B relative astrometry with a precision of approximately $234$~$\mu$as. The main technical result was that the complete acquisition, fringe-tracking, science-recording, and astrometric chain operated successfully on sky. The remaining tests are to check sensitivity limits: observing a faint secondary requires stable phase tracking and coherent integrations substantially longer than the usual tens of milliseconds.

\subsection{Science goals beyond the demonstration}

The targets accessible with this mode are mainly determined by the availability of a bright reference star, typically $H<7$, within the approximately 5 arcsec dual-field\cite{Anugu2025DualStar} interferometric field of view. A primary science goal is the detection of faint companions. A bright primary can provide phase referencing for a fainter visual secondary, enabling searches for close companions around the secondary and therefore for hierarchical triple systems. Multi-epoch differential astrometry can also refine the wide orbit and reveal short-period astrometric wobbles caused by an unresolved subsystem or exoplanet \cite{Gardner2022,Gravity2024}. Measurements of separation, near-infrared flux ratio, and orbital motion can then be combined with spectroscopy and Gaia astrometry to determine dynamical masses.

\subsection{Dual-field operation with all-in-one and ABCD beam combiners}

The CHARA dual-field mode is primarily a software-enabled capability that reuses existing hardware from CHARA, MIRC-X, MYSTIC, and the STST. Dual-star acquisition at the telescopes and adaptive-optics systems\cite{Che2013,Anugu2020AO} follows the standard acquisition procedure. After propagation through the CHARA delay lines, the two stars are imaged by the STST in the beam-combination laboratory. We developed software that uses the measured STST positions to optimize the injection of each star into the MIRC-X and MYSTIC single-mode fibers. Once both beams are injected, fringes are acquired on the phase reference beam combiner channel and stabilized. A separate dual-star fringe-acquisition routine uses the measured position of the companion relative to the primary to predict the differential delay and locate fringes on the science combiner.

Dual-field interferometry with MIRC-X and MYSTIC can be implemented using either all-in-one or pairwise ABCD beam combiners. Both instruments contain mature all-in-one (AIO) image-plane combiners. They also contain pairwise ABCD integrated-optics options. The SPICA-FT combiner installed in MIRC-X is a six-telescope, $H$-band integrated-optics combiner \cite{Pannetier2022}, while MYSTIC contains a four-telescope ABCD integrated-optics combiner, a spare from the GRAVITY project \cite{Setterholm2023MYSTIC}. Table~\ref{tab:combiners} summarizes the available configurations. Our dual-field experiments have so far used the AIO combiners, mainly because a complete offline data-reduction pipeline for the ABCD combiners is not yet operational, although the real-time acquisition software is available.

\begin{table}[t]
\caption{Beam-combination options relevant to dual-field operation.}
\label{tab:combiners}
\centering
\small
\begin{tabularx}{\linewidth}{p{4.5cm}p{0.5cm}p{0.5cm}p{4.0cm}Y}
\toprule
Combiner & Band & $N_{\rm tel}$ & Type & Present role and maturity \\
\midrule
MIRC-X AIO (PI: Monnier \& Kraus) & $J/H$ &  6 & Image-plane, all-in-one &
Workhorse science mode; phase tracking demonstrated in $J$ and $H$. Science data collection. \\
SPICA-FT (PI: Mourard) & $H$ &  6 & Pairwise ABCD, integrated optics & Phase tracking and science data collection. \\
MYSTIC AIO (PI: Monnier) & $K$ &  6 & Image-plane, all-in-one &
Workhorse science mode; phase tracking demonstrated in $K$. Science data collection. \\
MYSTIC ABCD (PI: Monnier) & $K$ & 4 & Pairwise ABCD, integrated optics &
Science data collection \\
\bottomrule
\end{tabularx}
\end{table}

\section{PHASE-TRACKING DEVELOPMENT IN THE $J$, $H$, AND $K$ BANDS}
\label{sec:tracking}

To freeze atmospheric piston over exposures longer than the atmospheric coherence time, phase-delay tracking is required in addition to the standard group-delay tracking used at CHARA. The SPICA-FT phase-tracking software \cite{Pannetier2022}, originally developed for SPICA observations, provides this capability for the dual-field mode. We have adapted the controller for $J$ and $H$-band tracking with MIRC-X and $K$-band tracking with MYSTIC.

The same controller architecture is used for both instruments. \\
Separate instances are run as 
\texttt{mircx\_opdcontroller\_server --mircx} or 
\texttt{mircx\_opdcontroller\_server --mystic}. The controller reads the coherent fluxes calculated by the MIRC-X or MYSTIC real-time software from shared memory, estimates the group and phase delays, and sends optical path difference (OPD) corrections to the CHARA delay lines \cite{Anugu2026JATIS}. To date, SPICA-FT has been operated primarily with the standard all-in-one combiners.

Initial on-sky tests give representative phase tracking residual OPD ranges of approximately 100--200~nm rms for $J$, $H$ and $K$. These values are preliminary performance ranges rather than universal error floors. The residuals vary among baselines and depend on target brightness, atmospheric seeing, and fringe-injection stability. A more complete characterization of SPICA-FT performance will be presented in forthcoming SPICA publications.


\section{EXPECTED SENSITIVITY FROM SIMULATIONS}
\label{sec:sensitivity}

\begin{figure}[t]
\centering
\includegraphics[width=0.72\textwidth]{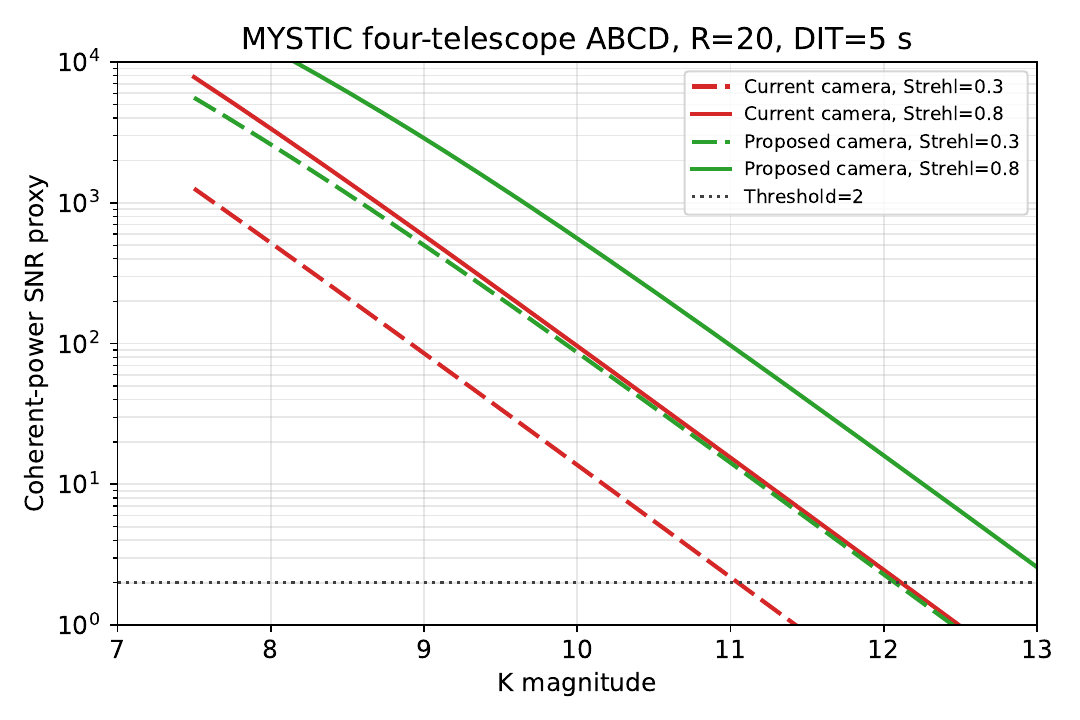}
\caption{Simulated coherent-power SNR for the four-telescope MYSTIC ABCD combiner at $R\simeq20$ with 5~s integrations. Red and green curves represent the current and proposed camera configurations, respectively; dashed and solid curves correspond to Strehl ratios of 0.3 and 0.8. The horizontal dotted line marks the threshold $Q_{\rm coh}=2$.}
\label{fig:sensitivity}
\end{figure}

Figure~\ref{fig:sensitivity} presents a preliminary $K$-band sensitivity simulation for the four-telescope MYSTIC ABCD combiner at $R\simeq20$, using 5~s detector integrations. The six baselines, four ABCD outputs, and six spectral channels correspond to $N_{\rm pix}=144$ detector pixels. The measured visibility is represented as
\begin{equation}
V_{\rm obs}=V_{\star}V_{\rm sys}V_{\rm atm},
\end{equation}
where $V_{\star}$ is the intrinsic source visibility and $V_{\rm sys}$ and $V_{\rm atm}$ represent instrumental and atmospheric visibility losses, respectively. We adopt $V_{\rm obs}^{2}=0.5$, representative of MYSTIC measurements of unresolved sources.

The detected source counts, total noise variance, coherent-fringe
signal-to-noise ratio (SNR), and coherent-power proxy are calculated as
\begin{align}
N_{\star} &=
F_{\rm ref}t\,10^{-0.4(K-K_{\rm ref})}
\left(\frac{\mathcal{S}}{\mathcal{S}_{\rm ref}}\right),\\
\sigma_{\rm tot}^{2} &=
N_{\star}+N_{\rm pix}
\left[(B_{\rm th}+B_{\rm dark})t+\sigma_{\rm R}^{2}\right],\\
\mathrm{SNR}_{\rm coh} &=
\frac{V_{\rm obs}N_{\star}}{\sigma_{\rm tot}},\\
Q_{\rm coh} &\equiv
\mathrm{SNR}_{\rm coh}^{2}
=
V_{\rm obs}^{2}\frac{N_{\star}^{2}}{\sigma_{\rm tot}^{2}}.
\end{align}

Here, $N_{\star}$ is the total detected source count during a coherent exposure $t$; $K$ is the target magnitude; $F_{\rm ref}$ is the total detected count rate at magnitude $K_{\rm ref}$ for phase tracking reference star and reference Strehl ratio $\mathcal{S}_{\rm ref}$; $B_{\rm th}$ and $B_{\rm dark}$ are the thermal-background and dark-current rates per pixel; $\sigma_{\rm R}$ is the read noise per pixel per frame; and $\sigma_{\rm tot}^{2}$ is the total variance.
Here, $\mathcal{S}$ is the science target Strehl ratio and
$\mathcal{S}_{\rm ref}$ is the phase reference star Strehl ratio.

We adopt $F_{\rm ref}=6.0\times10^{4}~e^-\mathrm{s}^{-1}$ across the 144 pixels at $K_{\rm ref}=6$ phase reference star (measured from previous data) and $\mathcal{S}_{\rm ref}=0.8$. Both camera configurations assume $\sigma_{\rm R}=1~e^-\mathrm{pixel}^{-1}\mathrm{frame}^{-1}$. The current configuration uses $B_{\rm th}=400~e^-\mathrm{pixel}^{-1}\mathrm{s}^{-1}$ and neglects dark current. The proposed configuration uses $B_{\rm th}=40~e^-\mathrm{pixel}^{-1}\mathrm{s}^{-1}$ and $B_{\rm dark}=20~e^-\mathrm{pixel}^{-1}\mathrm{s}^{-1}$.

For a threshold of $Q_{\rm coh}=2$, the predicted limits
range from approximately $K=11.1$ for the current camera at Strehl 0.3
to $K=13.1$ for the proposed configuration at Strehl 0.8. These are simulated rather than achieved on-sky limits.

\section{UPGRADES TOWARD FAINT DUAL-FIELD OPERATION}
\label{sec:upgrades}

Reaching faint targets requires three practical improvements: stable acquisition of the secondary, preservation of phase coherence during long integrations, and reduction of the thermal and detector background. The following planned upgrades address these
limitations.

\subsection{Six Telescope Star Tracker Upgrade}

The STST is used to acquire both stars, measure their relative field positions, and maintain their injection into the MIRC-X and MYSTIC fiber sets. It currently uses a C-RED~2 InGaAs camera whose read noise ($\sim30~e^-\mathrm{pixel}^{-1}$) and dark current ($\sim365~e^-\mathrm{pixel}^{-1}\mathrm{s}^{-1}$) limit acquisition to relatively bright targets, typically $H<6$. An NSF-supported upgrade led by G. Schaefer will replace this detector with a lower-noise C-RED ONE camera. The new configuration will also use previously unexploited wavelength ranges, receiving nearly all available light between 850 and 1000~nm and approximately 4\% between 1000 and 1650~nm. The design goal is to extend acquisition to approximately $H\sim13$. Faster operation should also improve faint-companion acquisition, provide more rapid correction of residual tip--tilt errors through the telescope adaptive-optics systems, stabilize fiber coupling, and reduce leakage from the primary star into the science channel.

\subsection{MYSTIC Thermal-Background Reduction}

MYSTIC is currently background limited during long $K$-band integrations. Its spectrograph cryostat and detector camera are separated by a warm window rather than sharing a continuous cold optical path. Thermal emission from this interface increases the detector background and can drive pixels toward saturation before the phase-tracking coherence time becomes the limiting factor. Removing the warm window and integrating the vacuum and cooling systems more closely have therefore been identified as important future MYSTIC upgrades. We are also seeking support for a new C-RED ONE camera with lower dark current and improved performance during long integrations.

\subsection{Telescope Adaptive-Optics Upgrade}

The achievable contrast in dual-field observations depends strongly on the Strehl ratio. A higher Strehl ratio improves coupling into the single-mode fibers and reduces leakage from the bright primary into the off-axis science channel. The current CHARA adaptive-optics systems provide H-band Strehl ratios of approximately 20--50\% \cite{Anugu2020AO}. A proposed upgrade aims to increase the near-infrared Strehl ratio to as high as approximately 80\%. Such an improvement would increase both fringe-tracker SNR and science-channel throughput while reducing primary-star contamination caused by imperfect fiber injection.

\section{SUMMARY AND NEXT STEPS}
\label{sec:summary}

CHARA has progressed from a dual-field concept to an on-sky demonstration. First-light with $\alpha$~Psc demonstrated simultaneous dual-star acquisition, fringe stabilization of the off-axis science fringes, detection of the Ba--Bb subsystem, and differential astrometry between components A and B. The next objective is to determine how far the science-channel sensitivity can be extended beyond the usual $H/K\simeq8$--9 limit. Observations for this sensitivity study have been obtained, and the long-exposure analysis is currently in progress.

The current system offers scientifically useful wavelength flexibility. MIRC-X can phase track in $J$ or $H$ while MYSTIC records long $K$-band integrations, or MYSTIC can track in $K$ while MIRC-X records $J/H$ science data. Both instruments provide all-in-one combiners, while pairwise ABCD options are available through SPICA-FT and the MYSTIC integrated-optics combiner. Phase tracking has been demonstrated in the $J$, $H$ and $K$ bands, with representative residuals of approximately 100--200~nm rms.

Sensitivity simulations indicate that stable 5~s coherent integrations could provide access to an approximate $K=11$--13 regime, depending strongly on Strehl ratio, detector performance, thermal background, and the adopted detection threshold. In parallel, the planned upgrades to the STST, the MYSTIC detector and cryogenic system, and CHARA adaptive optics provide a path toward reliable observations of faint off-axis targets. The complete long-exposure analysis and any companion-search results from the sensitivity-demonstration observations will be presented in a separate paper.

\acknowledgments
This work is based upon observations obtained with the Georgia State University Center for High Angular Resolution Astronomy Array at Mount Wilson Observatory.  The CHARA Array is supported by the National Science Foundation under Grant No. AST-2034336 and AST-2407956. Institutional support has been provided from the GSU College of Arts and Sciences, Office of the Provost, and Office of the Vice President for Research and Economic Development. SK acknowledges funding for MIRC-X from the European Research Council (ERC) under the European Union's Horizon 2020 research and innovation programme (Starting Grant No. 639889 and Consolidated Grant No. 101003096). JDM acknowledges funding for the development of MIRC-X (NASA-XRP NNX16AD43G, NSF-AST 2009489) and MYSTIC (NSF-ATI 1506540, NSF-AST 1909165). DM acknowledges funding for SPICA-VIS/SPICA-FT from the European
Research Council (ERC) under the European Union’s Horizon 2020 research and innovation programme (Grant agreement No. 101019653).

\bibliography{references}
\bibliographystyle{spiebib}

\end{document}